\pdfoutput=1
\documentclass[aps,prl,twocolumn,superscriptaddress,preprintnumbers,floatfix,nofootinbib,notitlepage,showkeys,showpacs]{revtex4-1}

\usepackage[utf8]{inputenc}
\usepackage{cancel}

\usepackage{graphicx}
\usepackage{hyperref}
\usepackage{latexsym}
\usepackage{amsmath}
\usepackage{amssymb}
\usepackage{bbm}
\usepackage{microtype}
\usepackage{nicefrac}

\usepackage[normalem]{ulem}
\usepackage{pdfsync}
\usepackage{epsfig}
\usepackage{epstopdf}
\usepackage{subfigure}
\usepackage{color}
\usepackage{comment}
\usepackage{slashed}
\usepackage{placeins}
\usepackage{cleveref}
\usepackage{xcolor}
\definecolor{cites}{RGB}{0,180,0}
\definecolor{links}{RGB}{200,0,0}
\hypersetup{colorlinks=true,citecolor=cites,linkcolor=links,urlcolor=blue}

\usepackage{graphicx}

\usepackage{tabu}
\usepackage{float}

\newcommand{\beq}{\begin{eqnarray}}
\newcommand{\eeq}{\end{eqnarray}}

\newcommand{\bmp}{\noindent\begin{minipage}{16cm}}
\newcommand{\emp}{\end{minipage}\vskip 7mm} 

    \newcommand{\dd}{\mathrm{d}}

    \newcommand{\bee}{\begin{equation}}
        \newcommand{\eee}{\end{equation}}

\def\lsim{\mathrel{\rlap{\lower4pt\hbox{\hskip1pt$\sim$}}
    \raise1pt\hbox{$<$}}}                
\def\gsim{\mathrel{\rlap{\lower4pt\hbox{\hskip1pt$\sim$}}
    \raise1pt\hbox{$>$}}}                

\usepackage{tikz}
\usetikzlibrary{arrows,shapes.misc,decorations.markings,decorations.pathmorphing,positioning,intersections,patterns}
\tikzset{cross/.style={cross out, draw=black, minimum size=2*(#1-\pgflinewidth), inner sep=0pt, outer sep=0pt},
cross/.default={1.5mm}}
\tikzset{mydash/.style={dashed, dash pattern=on 4pt off 5pt}}
\tikzset{
  vertex/.style={draw,shape=circle,fill=black,minimum size=3pt,inner sep=0pt},
  cross/.style={cross out, draw=black,thick, minimum size=6pt, inner sep=0pt, outer sep=0pt},
  cross2/.style={path picture={\draw[black](path picture bounding box.south east) -- (path picture bounding box.north west) (path picture bounding box.south west) -- (path picture bounding box.north east);
}}
  external/.style={inner sep=2pt},
  plabel/.style={inner sep=2pt},
  blob/.style={circle,minimum size=0.7cm,fill=black!20,draw,thick,pattern=north west lines, pattern color=black!20},
  dblob/.style={circle,size=5pt,fill=black!20,draw,thick,pattern=north west lines, pattern color=black!20},
  whiteblob/.style={circle,fill=white,minimum size=1.0cm,draw,thick},
  whiteblob2/.style={circle,fill=white,minimum size=1.0cm},
  effective/.style={rectangle,fill=black!20,minimum size=0.5cm,draw,thick},
  vev/.style={shape=vev,draw,inner sep=2pt,thick},
  mass/.style={shape=cross,draw,thick},
  rscalar/.style={dashed,thick},
  mfermion/.style={thick},
  scalar/.style={postaction={decorate}, decoration={markings,mark=at position .55 with {\arrow{latex}}},dashed,thick},
  ooscalar/.style={postaction={decorate}, decoration={markings,mark=at position .7 with {\arrow{latex}}},dashed,thick},
  fermion/.style={postaction={decorate}, decoration={markings,mark=at position .55 with {\arrow{latex}}},thick},
  majfermion/.style={postaction={decorate}, decoration={markings,mark=at position .7 with {\arrow{latex}}},thick},
  oofermion/.style={postaction={decorate}, decoration={markings,mark=at position .85 with {\arrow{latex}}, mark=at position .35 with {\arrowreversed{latex}}},thick},
  iifermion/.style={postaction={decorate}, decoration={markings,mark=at position .35 with {\arrowreversed{latex}}, mark=at position .85 with {\arrow{latex}}},thick},
  gaugeboson/.style={decorate, decoration={snake},thick},
  gluon/.style={decorate, decoration={coil,amplitude=4pt, segment length=5pt},thick},
  photon/.style={decorate, decoration={snake},thick},
  dashdot/.style={dash pattern=on .4pt off 3pt on 4pt off 3pt,thick}
}

\begin{document}
\title{A critical look at $\beta$-function singularities at large $N$}
\author{Tommi Alanne}
\email{tommi.alanne@mpi-hd.mpg.de}
\affiliation{Max-Planck-Institut f\"{u}r Kernphysik, 
    Saupfercheckweg 1, 69117 Heidelberg, Germany}
\author{Simone Blasi}
\email{simone.blasi@mpi-hd.mpg.de}
\affiliation{Max-Planck-Institut f\"{u}r Kernphysik, 
    Saupfercheckweg 1, 69117 Heidelberg, Germany}
\author{Nicola Andrea Dondi}
\email{dondi@cp3.sdu.dk}
\affiliation{CP$^3$-Origins, University of Southern Denmark, Campusvej 55, 5230 Odense M, Denmark}

\begin{abstract}
We propose a self-consistency equation for the $\beta$-function for theories with a large number of flavours, $N$, that
exploits all the available information in the Wilson--Fisher critical exponent, $\omega$, truncated at a fixed order in $1/N$.
We show that singularities appearing in critical exponents do not necessarily
imply singularities in the $\beta$-function.
We apply our method to (non-)abelian gauge theory, where $\omega$
features a negative singularity.
The singularities in the $\beta$-function and in the fermion mass
anomalous dimension are simultaneously removed providing no hint for a UV fixed point in the large-$N$ limit.

\end{abstract}

\preprint{CP3-Origins-2019-22 DNRF90}

\maketitle

\emph{Introduction.}---There are indications that perturbative series in quantum field theory are,
in general, asymptotic series with zero radius of convergence.
In theories with a large number of flavour-like degrees of freedom, $N$,
a re-organization of the perturbative expansion in powers of $1/N$
is convenient.
It can be shown that at fixed order in $1/N$ expansion,
the number of diagrams contributing grows only polynomially
rather than factorially:
convergent series are obtained that can be summed up within their radius of convergence.

There is a vast literature on resummed results corresponding to the first few orders in $1/N$ expansion, mainly for RG functions obtained via direct diagram resummation or critical-point methods,
see e.g. Refs~\cite{Espriu:1982pb,PalanquesMestre:1983zy,Kowalska:2017pkt,Antipin:2018zdg,Alanne:2018ene,Alanne:2018csn,Vasiliev:1981yc,Vasiliev:1981dg,Vasiliev:1982dc,Gracey:1993ua,Gracey:1996he,Ciuchini:1999wy,Gracey:1990wi,Gracey:1992cp,Derkachov:1993uw,Vasiliev:1992wr,Vasiliev:1993pi,Gracey:1993kb,Gracey:1993kc,Gracey:2017fzu,Manashov:2017rrx,Gracey:2018ame}.

Since the perturbative series at fixed order in $1/N$ are convergent,
singularities in the (generically complex) coupling are expected.
Appearance of such singularities on the real-coupling axis seems to be true for all the $d=4$ theories analyzed so far,
thereby having a dramatic effect on RG flows.
In particular, the appearance of singularities in the coefficients of the $1/N$ expansion for gauge and
Yukawa $\beta$-functions have inspired speculations of a possible UV fixed point~\cite{Mann:2017wzh,
Pelaggi:2017abg,Antipin:2017ebo,Molinaro:2018kjz,Cacciapaglia:2018avr,Sannino:2019sch,Cai:2019mtu}.

More generally, the UV fate of gauge theories for which asymptotic freedom is lost has
broad theoretical interest, and this is in fact the case of matter-dominated theories.
There, a non-trivial zero of the $\beta$-function can be envisaged if
the large-$N$ resummation produces a contribution to $\beta$ functions such that
${\lim_{g \rightarrow r} \beta^{1/N}(g) = - \infty,}$
where $r$ is the radius of convergence of the $1/N$ series.
Near the singularity, the $\mathcal{O}(1/N)$ contribution exceeds the leading-order result, and  it is
clear that a zero must emerge. Unfortunately, close to the radius of convergence the perturbative expansion in $1/N$
is broken, and higher-order
cannot be neglected.
Further shadow on the existence
of the fixed point as a consistent conformal field theory is cast by studying anomalous dimensions of other
operators in the vicinity of the $\beta$-function singularity:
in the case of large-$N$ QED truncated at $\mathcal{O}(1/N)$,
the anomalous dimension of the fermion mass diverges~\cite{Espriu:1982pb,PalanquesMestre:1983zy}, and
 it was recently pointed out that in the large-$N$ QCD the anomalous dimension
of the glueball operator breaks the unitarity bound~\cite{Ryttov:2019aux}.
Recently, the first lattice simulations
to investigate the existence of possible fixed points appeared~\cite{Leino:2019qwk}.
Even though these studies are not yet conclusive, no support for the fixed point is found.

In this letter we provide quantitative evidence that these
singularities are an artifact of the
fixed-order large-$N$ expansion of the $\beta$-function.
This follows from the observation that a fixed-order truncation
in $1/N$ in the critical
exponents is $\textit{not}$ equivalent to the same-order
truncation in $\beta$-functions, see also Ref.~\cite{Alanne:2019meg}.
Instead, a fixed-order critical exponent induces
higher-order terms in $1/N$.
These do not significantly affect the result far from the singular
point, but are relevant
near the singularity signaling a breakdown of the $1/N$ expansion.
In particular,
close to the radius of convergence these contributions diverge with alternating signs.
Remarkably, such contributions can be resummed, and the final result is free of singularities.
This conclusion is essential for the studies speculating on the UV fixed point, since they fully rely on the existence of a singularity in the $\beta$-function.

We demonstrate this method concretely for four-dimensional
gauge $\beta$-function and Gross--Neveu (GN)
model in two dimensions.
Generically, we find that the fixed-order
singularities are removed and the appearance of a fixed point
is not supported within the large-$N$ framework.

\emph{$\beta$-function from the critical exponents.}---Following
Ref.~\cite{Alanne:2019meg}, we review the general form for the $\beta$-function in the large-$N$
limit written in terms of the critical exponent, $\omega$. This critical exponent gives the slope of the $\beta$-function at the
Wilson--Fisher (WF) fixed point, $\beta(g_c)=0$,
\begin{equation}
    \label{eq:betaprime}
    \beta'(g_c)=\omega(d)\equiv -(d-d_c)+\sum_{n=1}^{\infty}\frac{\omega^{(n)}(d)}{N^n}\,,
\end{equation}
where $d$ is the dimension of spacetime\footnote{In the literature this equation is often found as $\omega = - \beta'/2$. We omit this factor for notational convenience.}.
The large-$N$ expansion of the $\beta$-function can be incorporated by using the
following ansatz:
\begin{equation}
    \label{eq:1couplbeta}
    \beta(g) = (d-d_c)g + g^2 \left( b N + c +
    \sum_{n=1}^\infty \frac{F_n(g N)}{N^{n-1}}\right),
\end{equation}
where $d_c$ is the critical dimension of the coupling $g$\footnote{
In QED, for example, $g=\alpha/\pi$, and should not be confused with the simple gauge coupling.},
$b$ and $c$ are model-dependent one-loop coefficients, and the functions $F_n$ satisfying $F_n(0) = 0$ are all-order
in $x\equiv gN$.

Using the ansatz one can relate the coupling value at the WF fixed point, $g_c$, to the spacetime dimension, $d$,
and, consequetively,
find the slope of the $\beta$ function, ${\beta'(g_c) = \omega( d)}$.

In Ref.~\cite{Alanne:2019meg}, we noticed that
the critical exponent $\omega^{(1)}$ contributes to
the $\beta$-function also beyond $\mathcal{O}(1/N)$.
Same holds for each $\omega^{(j)}$: it contributes to all $F_n$ with $n\geq j$.
In the following, we denote the contribution of
$\omega^{(1)},\dots,\omega^{(j)}$ to $F_n$, $n\geq j$, by $F_n^{(j)}$.
It is worth to stress that these contributions are necessary in order to obtain the correct perturbative result from the critical point formalism.
It is tempting to assume that these originate from a specific class of nested diagrams.

Since $\omega^{(1)}$, or equivalently $F_1$, is known, all the
$F_n^{(1)}$ can be computed. These induced coefficients are found in closed form as
\begin{align}\label{eq:Frec}
F_1^{(1)}(x) &= F_1(x) = \int_0^x \frac{\dd t}{t^2} \omega^{(1)}(d_c - b t) \\
F^{(1)}_{n>1}(x) &= \int_0^x \frac{\dd t}{t^2}
\sum_{\ell=1}^{n-1} \frac{1}{\ell!}c_{n-\ell-1}^{(\ell)}
\left( \frac{t}{b} \right)^{\ell} \frac{\dd^\ell}{\dd t^\ell} \left[
t^2 F_1'(t)\right],\label{eq:Fn1}
\end{align}
where the $c^{(k)}_m$ are defined iteratively:
\begin{align}
c_0^{(k)} &= (F_1+c)^k \\
c_n^{(k)} &= \frac{1}{n (F_1+c)} \sum_{q=1}^n (qk +q -n) F_{q+1} c^{(k)}_{n-q}.
\end{align}



\begin{figure}
    \begin{center}
	\includegraphics[width=0.45\textwidth]{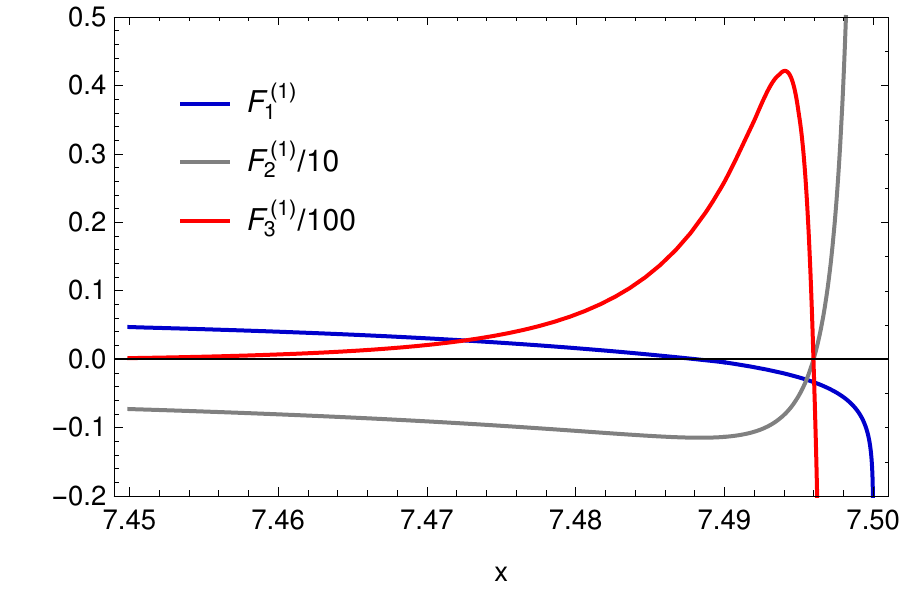}
    \end{center}
    \caption{The functions $F^{(1)}_{1,2,3}$ in the case of QED.}
    \label{fig:QEDF123}
\end{figure}

It follows from Eq.\,\eqref{eq:Fn1} that
if $F_1(x)$ features a negative singularity
at a given $x$, this results into sequence
of singularities of alternating
signs in $F_n^{(1)}$. A concrete example is given by
QED: we show $F_1^{(1)}$, $F_2^{(1)}$ and $F_3^{(1)}$ in Fig.~\ref{fig:QEDF123}.
This means that the negative pole in the $\beta$-function driven by
$F_1$ is not guaranteed
to persist when all the $F_n^{(1)}$ are taken into account.
In the next section, we show that all the $F_n^{(1)}$'s can be actually resummed, and the final result
features no singularity.

\emph{Self-consistency equation.}---A direct resummation of the $F_n^{(j)}$ terms
is not straightforward, and
we therefore employ a different approach. Denoting
\begin{equation}
\label{eq:Fcal}
 \mathcal{F}(x,N) \equiv \sum_{n=1}^\infty \frac{F_n(x)}{N^{n-1}}
\end{equation}
the relation
$\beta^\prime(g_c) = \omega(d)$ is rewritten as
\begin{equation}\label{eq:slope}
 -(d-d_c) + \frac{x_c^2}{N}\, \partial_x \mathcal{F}(x_c,N) = \omega(d),
\end{equation}
where the dimension and the critical coupling are related via
\begin{equation}
\label{eq:dim}
d=d_c - x_c \left(b + \frac{c + \mathcal{F}(x_c,N)}{N} \right)
\end{equation}

Equation~\eqref{eq:slope} would provide an exact solution,
if $\omega$ were known to all orders. However, in practice this is not
the case, but rather we have access to the contributions
induced by $\omega^{(1)},\dots,\omega^{(j)}$ only.
Nonetheless, a consistent solution to Eq.~\eqref{eq:slope}
incorporating all known coefficients can be
achieved by truncating the critical exponent to
\begin{equation}
 \omega(d) = - (d - d_c) + \sum_{n=1}^j\frac{1}{N^n} \omega^{(n)}(d),
\end{equation}
which corresponds to truncating $F_n$ to $F_n^{(j)}$ in $\mathcal{F}(x,N)$,
Eq.~\eqref{eq:Fcal}.
The resulting function is denoted by $\mathcal{F}^{(j)}(x,N)$.

Let us now concentrate on the simplest case $j=1$, where the truncation
leads to the following differential equation for $\mathcal{F}^{(1)}$:
\begin{equation}
\label{eq:Fp}
\begin{split}
 \partial_x \mathcal{F}^{(1)}(&x,N) =  \frac{1}{x^2} \omega^{(1)}(d) \\
 = & \frac{1}{x^2} \omega^{(1)}\left(
 d_c - x \left(b + \frac{c + \mathcal{F}^{(1)}(x,N)}{N} \right)\right)~,
 \end{split}
\end{equation}
where we have used Eq.\,\eqref{eq:dim}.
If the critical exponent as a function of space-time dimension is known, this is a non-linear first-order differential equation for $\mathcal{F}^{(1)}$.
Traditionally, this has been solved order by order in the $1/N$ expansion. 
Indeed, neglecting the back-reaction of $\mathcal{F}^{(1)}$ on the right-hand side of Eq.~\eqref{eq:Fp}
gives the standard solution ${\mathcal{F}^{(1)}(x,\infty) \equiv F_1(x)}$.
The advantage now is that we can solve Eq.\,\eqref{eq:Fp} as it is and
only afterwards take the large-$N$ limit.
This is equivalent to resumming all the
$F_n^{(1)}$'s, given explicitly in Eq.~\eqref{eq:Fn1}, that we know to be important near the singularity.

Where the $1/N$ expansion is under control,
the one-loop term in the $\beta$-function,
$g^2b N $, dominates and, in particular, no zero can emerge
for large enough $N$.
However, there exist examples in which
the critical exponent, $\omega^{(1)}$,
features a singularity for some real value of $d$,
potentially affecting the previous conclusion.
For instance,
in QED the first singularity of $\omega^{(1)}_{\mathrm{QED}}$ occurs at ${d=-1}$
translating to the $\mathcal{O}(1/N)$
singularity of the $\beta$-function at $x=7.5$.

Let us first consider a model where $b$ and the singularity in $\omega$ are of same sign: the
higher-order terms would just enhance the singularity
and lead to a Landau pole as is the case of
super-QED at $\mathcal{O}(1/N)$~\cite{Ferreira:1997bi} and
in $\mathrm{O}(N)$ model at $\mathcal{O}(1/N^2)$~\cite{Gracey:1996ub}.

On the contrary,
if the singularity and $b$ are of opposite sign,
as in QED and QCD, Eq.~\eqref{eq:Fp}
yields a smooth solution
which, close to the would-be-singularity at
$x = x_s$, approaches a \emph{scaling solution}
of the form:
\begin{equation}\label{eq:f}
 \begin{split}
& \mathcal{F}^{(1)}(x,N) = N \left(\frac{a}{x} -b \right) - c,
\quad x \gtrsim x_s,
 \end{split}
\end{equation}
where $a$ is typically $\mathcal{O}(1)$
and implicitly defined by
\begin{equation}\label{eq:a}
 a N = - \omega^{(1)}(d_c - a).
\end{equation}
This indicates that the alternating
singularities in the $F_n^{(1)}$ can be resummed
to yield a finite contribution.
By using Eq.~\eqref{eq:f}
and recalling that $ x = g N$, the $\beta$ function
is found to be
\begin{equation}\label{eq:scalbeta}
 \beta(g) 
 = a g \quad g \gtrsim g_s.
\end{equation}
Given that $a$ and $b$ need to have the same sign
due to the boundary condition
$\mathcal{F}(0,N) = 0$ and $\omega(d_c) = 0$,
a zero cannot emerge neither for $g \gtrsim g_s$,
nor for $g < g_s$, where the one-loop
coefficient dominates.

When the $\mathcal{O}(1/N^2)$ term, $\omega^{(2)}$,
 is included in the analysis, there are two possibilites:
\begin{enumerate}
 \item the closest singularity at $x = x_s^{(2)}$ is positive,
 \item the closest singularity at $x = x_s^{(2)}$ is negative.
\end{enumerate}

In the first case, the $\beta$-function clearly grows faster
than before close to $x_s^{(2)}$, and no zero can appear.
If the new singularity
is closer, this rather implies a Landau pole.
As for the regular points before the first singularity,
the contribution of $\omega^{(2)}$ is negligible for large enough $N$.
An example of this behaviour is
given by the $\mathrm{O}(N)$ model~\cite{Gracey:1996ub}.

In the second case, the same reasoning resumming
the alternating singularities applies
and gives the asymptotic scaling in Eq.~\eqref{eq:scalbeta}
with a modified coefficient $a$,
\begin{equation}\label{eq:aN}
 a = - \frac{1}{N} \omega^{(1)}(d_c - a) - \frac{1}{N^2}\omega^{(2)}(d_c - a),
\end{equation}
valid for $g \gtrsim \text{min}(g_s,g_s^{(2)})$. The same procedure generalizes to any finite order $\omega^{(j)}$.

To summarize, the singularities appearing in fixed-order critical exponents do not necessarily
imply singularities in the $\beta$-function. In particular no hint for a UV zero is found in the large-$N$ limit, as its existence relied entirely on the presence of a singularity.

Finally, we emphasize that the resummation we have employed is relevant
also beyond the case when the $\beta$-function features singularities
on the positive real axis. In the
following we will show that the wild
oscillations in the $\beta$-function of the Gross--Neveu (GN) model---which naively would lead
to infinitely many alternating IR and UV zeroes---can be resummed
in the same way.

As explicit examples we consider
two classes of models: four-dimensional gauge theories and GN model in two dimensions.
For the latter,
the critical exponent
is known up to $\mathcal{O}(1/N^2)$ allowing us to study the effect of higher-order corrections.

\begin{figure}[t]
\centering
\includegraphics[width=0.45\textwidth]{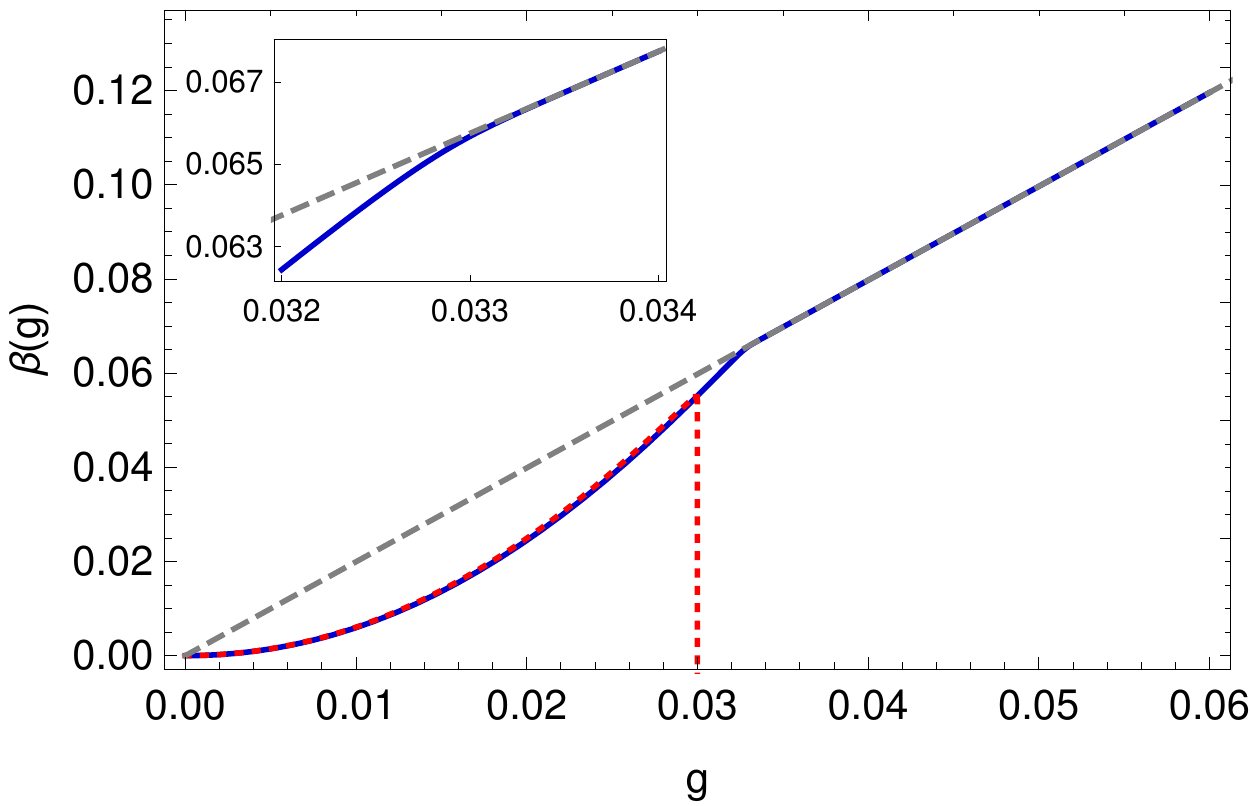}
\caption{The $\beta$-function for
QCD for $N = 100$ computed
numerically according to Eq.~\eqref{eq:Fp}. Dashed
line indicates the scaling solution.
The dotted line shows the singular solution one would encounter
neglecting the back-reaction of
$\mathcal{F}^{(1)}$ on the right-hand side of Eq.~\eqref{eq:Fp}.
}
\label{fig:pQ}
\end{figure}

\emph{QED \& QCD.}---The critical exponent for a general gauge $\beta$-function is known up to $\mathcal{O}(1/N)$
and is given in $d=2\mu$ by~\cite{Gracey:1996he}
\begin{align}
    \label{eq:omGauge}
    \omega^{(1)}(2\mu)=&\frac{\eta^{(1)}(2\mu)}{T_F}\left(\vphantom{\frac12}(2\mu-3)(\mu-3)C_F\right.\\
    &\left. -\frac{(4\mu^4-18\mu^3+44\mu^2-45\mu+14)C_A}{4(2\mu-1)(\mu)}\right),\nonumber
\end{align}
where $T_F$ and $C_F$ are the index and quadratic Casimir of the fermion representation, resp., $C_A$ is the Casimir of the adjoint representation, and
$\eta^{(1)}$ reads
\begin{equation}
    \label{eq:etaGauge}
    \begin{split}
	\eta^{(1)}(2\mu)&=\frac{(2\mu-1)(\mu-2)\Gamma(2\mu)}{4\Gamma(\mu)^2\Gamma(\mu+1)\Gamma(2-\mu)}.
    \end{split}
\end{equation}

For the
abelian case, the first singularity occurs at $\mu=-1/2$, while the non-abelian system has a singularity already at $\mu=1$.

We compute the $\beta$-function by solving Eq.~\eqref{eq:Fp} numerically for a benchmark value $N=100$.
In the notation of Eq.~\eqref{eq:Fp},
QED corresponds to $b=2/3$, $c=0$, while QCD is characterised by
$b=2/3, c=-11$. The scaling solutions are given by $a_{\mathrm{QED}}\approx4.995$,
$a_{\mathrm{QCD}}\approx 1.985$. In Fig.~\ref{fig:pQ} we show the numerical solution to Eq.~\eqref{eq:Fp}
for
QCD with $N=100$; for QED the plot looks qualitatively the same.
As expected from the general analysis above, the singularities and the putative UV fixed points at $x=3$ for QCD and $x=7.5$ for
QED have both diappeared.

\begin{figure}
\includegraphics[width=0.45\textwidth]{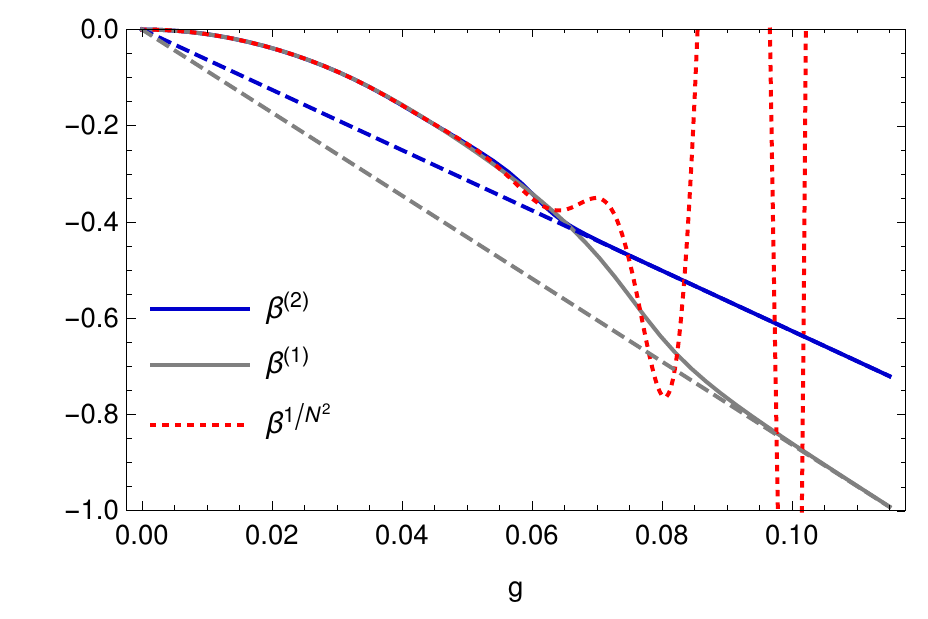}
    \caption{The solid lines show the GN $\beta$-function $\beta^{(2)}$ ($\beta^{(1)}$) for $N = 100$ computed numerically
    according to Eq.~\eqref{eq:Fp} using $\mathcal{F}^{(2)}$ ($\mathcal{F}^{(1)}$), and the
	dashed lines indicate the corresponding scaling solutions.
	The dotted red line depicts the $\mathcal{O}(1/N^2)$
	$\beta$-function without
	resummation. The scaling solution using $\lambda^{(1)}$ only is given
	by $a^{(1)}\approx -8.6$, while including $\lambda^{(2)}$ modifies this to $a^{(2)}\approx-6.3$.}
    \label{fig:GN}
\end{figure}

In the QED case, the fermion mass anomalous dimension has a singularity at the same coupling value
as the first singularity of $\omega^{(1)}$, $x=7.5$.
A fixed point in this coupling region would have the operator $\bar{\psi}\psi$ violating the unitarity bound.
Similarly as the critical exponent, $\omega$, we
truncate the fermion mass anomalous dimension to $\mathcal{O}(1/N)$:
\begin{equation}
    \label{eq:gammaMtru}
    \gamma_m=\frac{\gamma_m^{(1)}(d)}{N}=\frac{\gamma_m^{(1)}
	\left[d_c - x \left(b + \frac{c + \mathcal{F}^{(1)}(x,N)}{N}\right) \right]}{N},
\end{equation}
where the $\mathcal{O}(1/N)$ result is given by
${\gamma_m^{(1)}(2\mu)=-2\eta^{(1)}(2\mu)/(\mu-2)}$~\cite{Gracey:1993ua}.
Evaluating Eq.~\eqref{eq:gammaMtru}
with the solution for $\mathcal{F}^{(1)}$, we obtain
$\gamma_m$ in the
same truncation as the $\beta$-function.
We find that the singularity in $\gamma_m$ is also
removed, and the anomalous dimension reaches a constant value above $x=7.5$
given by
\begin{equation}
 \tilde{\gamma}_m = \frac{1}{N} \gamma_m^{(1)}(d_c - a_{\mathrm{QED}}).
\end{equation}
For $N=100$, we find $\tilde{\gamma}_m\approx -0.14$.

\emph{Gross--Neveu model.}---
The critical exponent, $\lambda(d) = \beta^{\prime}(g_c)$,
for the GN model is currently known up to $\mathcal{O}(1/N^2)$~\cite{Gracey:1993kb}.
The $\mathcal{O}(1/N)$ coefficient is explicitly given by
\begin{equation}
    \label{eq:}
    \lambda^{(1)}(2\mu)=\frac{4(\mu-1)^2\Gamma(2\mu)}{\Gamma(2-\mu)\Gamma(\mu)^2\Gamma(\mu+1)},
\end{equation}
while the expression for $\lambda_2(d)$ can be explicitly found in Ref.~\cite{Gracey:1993kb}.

In the notations of Eq.~\eqref{eq:1couplbeta}, the GN model is
characterized by $d_c = 2$, $b=-1$ and $c = 2$. We solve again Eq.~\eqref{eq:Fp} numerically for benchmark
value $N=100$ both using only the $\mathcal{O}(1/N)$ and $\mathcal{O}(1/N^2)$ critical exponent, $\lambda$.
We show the resulting $\beta$-functions in Fig.~\ref{fig:GN} along with the $\beta$-function computed
directly up to $\mathcal{O}(1/N^2)$ using Eq.~\eqref{eq:1couplbeta}.
The scaling solution using only $\lambda^{(1)}$ is given by $a^{(1)}\approx -8.6$,
while including $\lambda^{(2)}$ modifies this to $a^{(2)}\approx-6.3$.
The result shows no hint for an IR fixed point in agreement with previous studies~\cite{Schonfeld:1975us,Choi:2016sxt}.

\FloatBarrier

\emph{Conclusions.}---We have shown that singularities in a fixed-order large-$N$ critical exponent do not necessarily
imply singularities in the
$\beta$-function.
This is due to the fact that a fixed-order critical exponent generates contributions to every subsequent order in $1/N$ in the $\beta$-function.
We proposed a self-consistency equation to properly include these contributions.

In the case of negative singularities that have inspired speculations of UV fixed points,
it turns out that the same singularity appears with alternating sign at higher-order terms, and resumming these contributions yields an asymptotic linear growth of the $\beta$-function rather than a UV zero.
As concrete examples we showed this scaling behavior in the case of QED, QCD and the GN model. For QED and QCD, the
singularities are removed and
in the GN model the wild oscillations tamed.
For QED, this procedure simultaneously cures the singularity of the fermion mass anomalous dimension.

We stress that the linear scaling is very sensitive to the higher-order corrections in $1/N$ to the critical exponent that could potentially
turn it into a Landau-pole behaviour. Nonetheless, the emergence of a fixed point remains incompatible within any finite set of higher-order corrections.
Our result invalidates fixed points based on the singularities in the large-$N$ $\beta$-function. A hypothetical fixed point could thus be supported only through
a non-perturbative computation beyond the $1/N$ expansion.
\\

We thank John Gracey  for valuable comments
and are grateful for the numerous fruitful discussions
during the MASS2019 conference. The CP$^3$-Origins centre
is partially funded by the Danish National Research Foundation, grant number DNRF:90.

\bibliography{refs.bib}
\onecolumngrid
\twocolumngrid

\end{document}